\documentclass[aps,floatfix,prl,superscriptaddress,groupedaddress]{revtex4}
\usepackage{graphicx,amsmath,bm}
\usepackage[caption=false]{subfig}
\usepackage{color}

\newcommand{\unitvec}[1]{\hat{\vec{#1}}}
\renewcommand{\vec}[1]{\bm{#1}}

\begin{document}

\title{Nonlocal Electromagnetic Response of Graphene Nanostructures}

\author{{ Arya Fallahi$^{1,*}$, Tony Low$^{2,3,4,*}$, Michele Tamagnone$^5$ and Julien Perruisseau-Carrier$^5$} \\
{\small \em $^1$ DESY-Center for Free-Electron Laser Science, Notkestrasse 85, D-22607 Hamburg, Germany \\
$^2$ Department of Physics and Electrical Engineering, Columbia University, New York, NY 10027, USA\\
$^3$ IBM T.J. Watson Research Center, Yorktown Heights, NY 10598, USA \\
$^4$ Department of Electrical and Computer Engineering, University of Minnesota, Minneapolis, MN 55455, USA \\
$^5$ Adaptive MicroNano Wave Systems, Ecole Polytechnique F\'{e}d\'{e}rale de Lausanne (EPFL), CH-1015 Lausanne, Switzerland }}

\date{\today}

\begin{abstract}
Nonlocal electromagnetic effects of graphene arise from its naturally dispersive dielectric response.
We present semi-analytical solutions of nonlocal Maxwell's equations for graphene nano-ribbons array with features around $100\,$nm, where we found prominent departures from its local response.
Interestingly, the nonlocal corrections are stronger for light polarization parallel to the ribbons, which manifests as additional broadening of the Drude peak. 
For the perpendicular polarization case, nonlocal effects lead to blue-shifts of the plasmon peaks. These manifestations provide a physical measure of nonlocal effects, and we quantify their dependence on ribbon width, doping and wavelength.
\end{abstract}

\maketitle

Constitutive relationships present equations that bridge the materials' properties of a medium to the electromagnetic fields within. 
In classical electromagnetism dating from the nineteenth century, methods for considering more complex material properties, such as anisotropy and non-linearity, have been developed and widely used \cite{jacksonBook}. 
In developing these relationships, nonlocal effects or so-called \emph{spatial dispersion} are neglected for most natural media \cite{landauBook,haleviBook}. 
However, the advances in nanotechnology and the capabilities for fabricating materials with controlled nanoscale geometries has enabled artificially structured metamaterials with features orders of magnitude smaller than the free space wavelength. 
In these regime, nonlocal effects are becoming increasingly important \cite{nonlocalMetamaterial,nonLocalEffectsTransformationOptics,infraredHologram}. 
In metallic devices, it was evidenced that structures with apex features lead to intense accumulation of electrons which in turn yields large nonlocal effects \cite{nonlocalMetal1,nonlocalMetal2,nonlocalMetal3}. 
In addition, it is shown that the propagation of surface plasmons are also influenced by the nonlocal effects \cite{nonlocalPlasmons}.
In these situations, one needs to solve Maxwell's equations for a medium in which the dielectric displacement vector $\vec{D}(\vec{r},\omega)$ depends not only on the electric field intensity $\vec{E}(\vec{r},\omega)$ but also on its spatial derivatives. \footnotetext[1]{Corresponding authors: Arya Fallahi (arya.fallahi@cfel.de) and Tony Low (tonyaslow@gmail.com)} 

In this letter, we discuss the nonlocal electromagnetic response in graphene, whose influence on the optical properties of graphene plasmons and nanostructures remains relatively unexplored \cite{gateTuningGraphene,graphenePlasmonicsNature,graphenePlasmonicsReview,graphenePlasmonicsNanoLetter,graphenePlasmonics,graphenePlasmonicsNanoImaging,grapheneDampingPathways,patternedGrapheneNature,grapheneStripes2,nonlocalGraphenePlasmons,grapheneSHG}. 
Recently, graphene received considerable attention as a plasmonic material \cite{gateTuningGraphene} due to its semi-metallic nature, allowing for electrically tunable plasmonic devices for terahertz to mid-infrared applications \cite{graphenePlasmonicsNature,graphenePlasmonicsReview,graphenePlasmonicsNanoLetter,graphenePlasmonics,graphenePlasmonicsNanoImaging}. 
The plasmonic resonances in graphene can be engineered by patterning it into nano-structures such as ribbons, disks, antidots, or stacks \cite{grapheneDampingPathways,patternedGrapheneNature,stripGrapheneNature}. 
In the terahertz to midinfrared spectral range, light can be confined to a volume a million times smaller than the diffraction limit, where nonlocal effects might be important, especially with the continual downscaling in sizes.

\emph{Model---} Graphene is traditionally modeled by a homogeneous local conductivity $\sigma(\omega)$ relating the current flowing on the surface $\vec{J}(\vec{r},\omega)$ to the imposed tangential electric field $\vec{E}(\vec{r},\omega)$. 
However, when taking the nonlocal effects into account, the surface current is both a function of the imposed electric field as well as its spatial derivatives. 
Written in the most general way for a linear 2D material, it bears the following form,
\begin{equation}
\vec{J}(\vec{r},\omega) = \sigma(\mathbf{\nabla},\omega) \vec{E}(\vec{r},\omega) = \left( \begin{array}{cc} \sigma_{xx} & \sigma_{xy} \\ \sigma_{yx} & \sigma_{yy} \end{array} \right)_{(\mathbf{\nabla},\omega)} \vec{E}(\vec{r},\omega),
\label{constituveRelationSpatial}
\end{equation}
where $\sigma$ is the graphene's conductivity tensor and $\mathbf{\nabla}=\unitvec{x} \partial / \partial x + \unitvec{y} \partial / \partial y$ denotes the spatial derivative operator on a surface. 
Coordinate transformation from spatial to spectral domain makes the sophisticated nonlocal equations tractable by transforming the spatial derivative operator into simple multiplications by the wave vector $\vec{k}$,
\begin{equation}
\tilde{\vec{J}}(\vec{k},\omega) = \sigma(\vec{k},\omega) \tilde{\vec{E}}(\vec{k},\omega),
\label{constituveRelationSpectral}
\end{equation}
where the tilde sign denotes quantities in the spectral representation.
Graphene's conductivity tensor $\sigma(\vec{k},\omega)$ can be obtained by various techniques \cite{grapheneConductivityModel,grapheneRPA1,grapheneRPA2,grapheneSDModels,grapheneSDSpaceTime,grapheneModeling3}. 
In this work, we employed the Bhatnagar-Gross-Krook (BGK) model \cite{grapheneSDModels}, which provides us with an analytical form for the longitudinal and azimuthal components of $\sigma(\vec{k},\omega)$.  
The BGK model compares well with the Kubo formula \cite{grapheneConductivityModel,grapheneRPA1,grapheneRPA2} over the near infrared frequencies, i.e. the spectral range used in this work, see Suppl. Info. 

From the theory of Green's functions, the radiated electric field in response to a current excitation $\vec{J}(\vec{r})$ can be written as,
\begin{equation}
\vec{E}(\vec{r}) = \int \vec{G}(\vec{r}-\vec{r}') \vec{J}(\vec{r}') d\vec{r}'.
\label{spatialGreenFunction}
\end{equation}
Introducing the coordinate transformation in spectral domain $ \tilde{\vec{\zeta}}(\vec{k}) = \int \vec{\zeta}(\vec{r}) \exp(-i \vec{k} \cdot \vec{r}) d\vec{r} $ (with the implicit $\exp(-i \omega t)$ time-harmonic dependence), the equation transforms to the following algebraic form,
\begin{equation}
\tilde{\vec{E}}(\vec{k}) = \tilde{\vec{G}}(\vec{k}) \tilde{\vec{J}}(\vec{k}).
\label{spectralGreenFunction}
\end{equation}
The superposition of the radiated field obtained from (\ref{spectralGreenFunction}) with the exciting field should satisfy the boundary condition at the graphene surface, which is formulated as
\begin{equation}
\tilde{\vec{E}}^{\mathrm{inc}}(\vec{k}) = \left( \tilde{\vec{G}}(\vec{k}) + Z(\vec{k}) \right) \tilde{\vec{J}}(\vec{k}),
\label{boundaryCondition}
\end{equation}
where $Z(\vec{k})$ is the surface impedance of the graphene layer, which by definition is just the inverse of the graphene's conductivity. 

In order to solve Maxwell's equation with the nonlocal graphene conductivity $\sigma(\vec{k},\omega)$ described above, we follow the method of moments (MoM) in the spectral domain \cite{periodicGrapheneMetasurfaces,grapheneGFR}. 
Here, we briefly outline the numerical procedure. 
In the spectral MoM, we first express the incident electric field in the spectral domain, i.e. $\tilde{\vec{E}}^{\mathrm{inc}}(\vec{k})$, using a Fourier transformation, which follows with evaluating the induced currents $\tilde{\vec{J}}(\vec{k})$ using (\ref{boundaryCondition}), and then computes the scattered field $\tilde{\vec{E}}(\vec{k})$ in spectral domain using (\ref{spectralGreenFunction}) \cite{periodicGrapheneMetasurfaces}. 
In order to account for the specific boundary conditions imposed on $\tilde{\vec{J}}(\vec{k})$, it is expanded in terms of well-suited basis functions. 
Consideration of nonlocal effects within this method is then straightforward.  
Instead of a constant surface impedance ($Z$), one employs a $\vec{k}$-dependent function for the corresponding spectral component of the impedance, i.e. $Z(\vec{k})$.
This procedure provides the unique possibility to account for nonlocal effects in graphene within the Maxwell's equations in a numerically tractable fashion.

\begin{figure}
\includegraphics[width=3.4in]{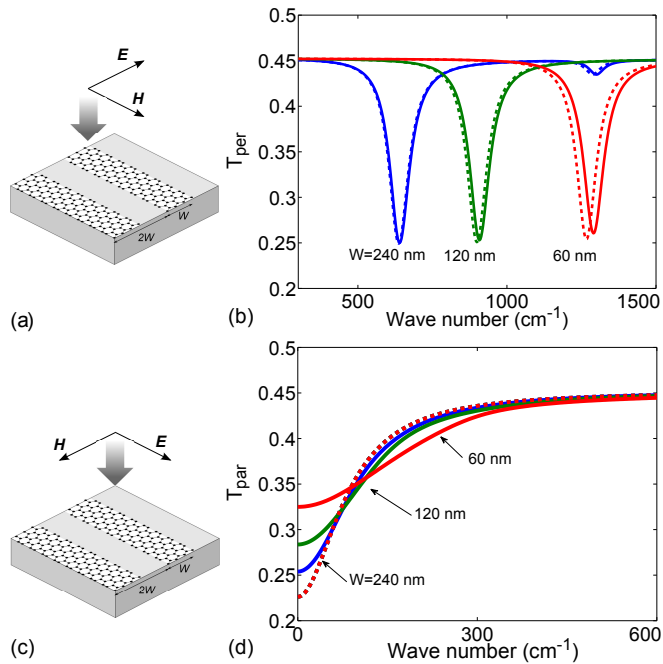}
\caption{Graphene nanoribbons array, doped at $\mu=0.5\,\mathrm{eV}$, residing on an infinitely thick substrate with light incidence at \textbf{(a)} perpendicular and \textbf{(c)} parallel polarization. Power transmission spectra are depicted in \textbf{(b)} and \textbf{(d)} respectively. The simulations are performed with (solid lines) and without (dashed lines) the spatial dispersion effects.}
\label{infiniteSubstrate}
\end{figure}

\emph{Results---} Using the proposed method, we investigate the interaction of an electromagnetic wave with graphene nanostructures. 
We consider the simplest setup where spatial dispersive effects can manifest, i.e. plane wave illumination on periodic ribbons array, residing on an infinitely thick substrate as illustrated in Fig.\,\ref{infiniteSubstrate}a and c. 
The inherent structural periodicity means that the problem can be formulated in discrete Fourier space, removing the requirement for any artificial discretization and the potential appearance of spurious numerical effects.
We assume a SiO$_2$ substrate with a dielectric constant $\epsilon_r = 3.9$. 
The simulations are carried out for different nano-ribbons widths, assuming filling factor of $\tfrac{1}{2}$, i.e. lattice constant of array twice the ribbon width. 
In the graphene's conductivity model, we have adopted typical experimental parameters \cite{grapheneDampingPathways}; carrier lifetime $\tau = 90\,\mathrm{fs}$, doping at $\mu=0.5\,\mathrm{eV}$, and an ambient temperature $T=300\,\mathrm{K}$. 

Fig.\,\ref{infiniteSubstrate}b and d show the computed transmission spectra for perpendicular ($T_{per}$) and parallel polarization ($T_{par}$), with and without spatial dispersion effects.
The resonance in the transmission spectra for perpendicular polarization is due to the excitation of localized plasmon modes \cite{grapheneDampingPathways,graphenePlasmonicsReview,graphenePlasmonicsEdgeWaveguide}. 
Spatial dispersion introduces a blue-shift of the resonant structure. 
Owing to the uniformity of the problem along the ribbons ($y$ direction), only the $q_x$-dependent components in the conductivity tensor are responsible for the observed nonlocal effects. 
Moreover, the electric field polarization would allow coupling only to the $\sigma_{xx}$ term.
To first order in the field variations in the graphene sheet (i.e. $\nabla \cdot \vec{E}$), $\sigma_{xx}$ has the following form (see also Suppl. Info. on full expression) \cite{grapheneSDModels},
\begin{equation}
\sigma_{xx} = \sigma_0 \left( 1 + v_F^2 \frac{3\omega+2i/\tau}{4\omega(\omega+i/\tau)^2} q_x^2 \right) \equiv \sigma_0 ( 1 + \alpha_{xx}^{\mathrm{SD}} ).
\label{conductivityFirstOrder}
\end{equation}
where $\sigma_0$ is the graphene's conductivity in the local limit. 
Hence, the dimensionless quantity $\alpha_{xx}^{\mathrm{SD}}\sim 3v_F^2q_x^2/(4\omega^2)$ dictates the amount of nonlocal effects.
In 2D graphene, the plasmon resonance is directly related to its conductivity via $\omega_{pl}= q_x\sigma_{xx}/2i\epsilon$, where $\epsilon$ is the effective dielectric permittivity of the two half spaces. 
In the local limit, it reduces to $\omega_{pl}\rightarrow\omega_{0}\equiv q_x\sigma_{0}/2i\epsilon$. 
The blue-shift due to spatial dispersion then bears the following form, 
\begin{equation}
\frac{\delta\omega_{pl}}{\omega_{pl}}\approx \frac{3v_f^2 \epsilon \pi\hbar^2}{2e^2}\frac{q_x}{\mu} \propto \frac{q_x}{k_f}
\label{blueshift}
\end{equation}
where $k_f$ is the Fermi wave-vector. 
The plasmon wave-vector is related to the ribbon width $W$ via, $q_x\approx 3\pi/4W$, after accounting for the anomalous reflection phase off the edges \cite{graphenePlasmonicsAnomalousReflection}. In our case, the proportionality constant in Eq.\,\ref{blueshift} is $\sim 0.4$.
The crossover between local to nonlocal electromagnetic response occurs when $q_x\sim k_f$, which unlike metals, can be tuned in graphene through doping.  
We defer this aspect to later discussion.

In the parallel polarization, light does not couple to plasmon, and the spectra resembles that of 2D graphene with a Drude peak at terahertz frequencies. Fig.\,\ref{infiniteSubstrate}d plots the transmission spectra for nanoribbon arrays with different $W$. 
Within a local conductivity model, the spectra produced all collapse together, since the induced current is directed along the ribbons and unable to distinguish the presence of the ribbon edges. 
Rather unexpectedly, a considerable change in the spectra is observed when spatial dispersion effects are included. 
In general, the nonlocal longitudinal conductivity can include terms quadratic in $q_x$ and $q_y$, and beyond \cite{grapheneSDModels}, where the induced currents picks up the spatial variations transverse to the electric field. 
From our calculations, the transmission in the $dc$ limit differs as much as $50\%$ in the $W=60\,$nm array. 
Nonlocal effects also produce an apparent broadening of the Drude peak.

\begin{figure}
\centering
\includegraphics[width=3.4in]{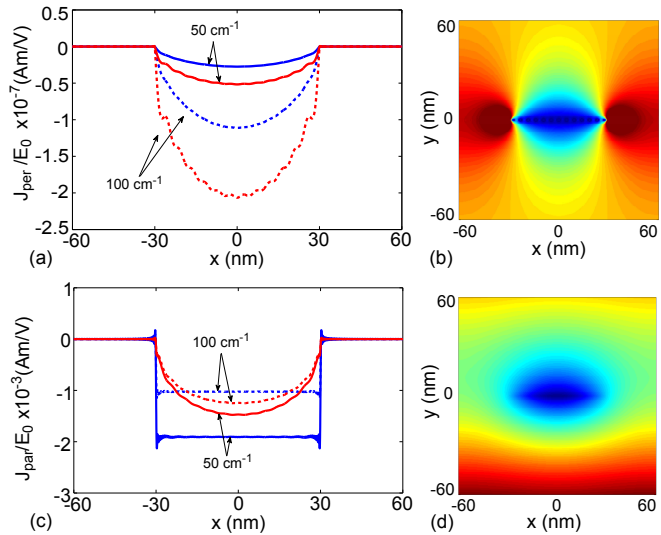}
\caption{Induced longitudinal currents in graphene nano-ribbons array ($W=60\,\mathrm{nm}$) with \textbf{(a)} perpendicularly and \textbf{(c)} parallelly polarized light at normal incidence. Calculations are performed for two frequencies as indicated, with (red lines) and without (blue lines) spatial dispersion. The transverse electric field component are displayed: $E_x$ for the \textbf{(b)} perpendicular (red for 1.0 and blue for 0.0) and $E_z$ for the \textbf{(d)} parallel polarization (red for 0.5957 and blue for 0.5956). All displayed results are normalized to the incident electric field $E_0$.}
\label{currentProfiles}
\end{figure}

Visualizing the induced currents in graphene can help to shed light on the disparate nonlocal response in these two polarizations.
Fig.\,\ref{currentProfiles} plots the induced longitudinal current and the radiated electric field for the $W=60\,\mathrm{nm}$ ribbons for perpendicular and parallel polarization respectively, at two terahertz frequencies i.e. $50\,$cm$^{-1}$ and $100\,$cm$^{-1}$. 
First, we see that the induced current in the parallel polarization is $4$ orders of magnitude larger than the perpendicular case. 
This is reasonable since the current in the latter is directed towards the ribbon edges. 
The induced current within the ribbon is very minimal, since the field varies only $1/1000$ of the free space wavelength across the ribbon, and has to be zero at the edges. 
This explains the larger induced current in the parallel polarization at terahertz.

Second, we observe an opposite dependence of the induced current with frequency for the two polarization. 
In the perpendicular polarization case, the current and its variations are larger at higher frequencies as depicted in Fig.\,\ref{currentProfiles}a. 
This follows from the argument above and (\ref{conductivityFirstOrder}). 
The parallel polarization case shows the opposite behavior. 
Fig.\,\ref{currentProfiles} illustrates the normalized transverse electric fields profile across the cross-section of the $60\,$nm ribbon for the two polarizations at $50\,$cm$^{-1}$. 
As shown in Fig.\,\ref{currentProfiles}d, the relative variation in $E$ is only around 0.01\% in the parallel polarization. 
However, since this variation occurs over the width of the ribbon, which is $1/1000$ of the free-space wavelength, it can translate to a large field gradient. 
We estimate that this electric field gradient can amount to equivalent conductivity variations with the same order of magnitude as the non-dispersive conductivity. 
Note that in this particular case, using a simple model with accuracy up to $q_x^2$ will not suffice. 
The nonlocal corrections in the conductivity can always be written in terms of $q/\omega$. 
Nonlocal effects are prominent in low frequency regime, resulting in large values for $q/\omega$ term. 
Therefore, unlike the perpendicular polarization case, nonlinear contributions from the higher order terms in the nonlocal conductivity become significant making the situation more complex and the development of a simple model very challenging. 
The equivalent conductivity variations, which is also spatially inhomogeneous, can effectively amount to a broadening in the spectral information. 
In our case, it manifests in an apparent broadening of the Drude peak in Fig.\,\ref{infiniteSubstrate}d. 
The spatial field variations also impose a qualitatively similar spatial profile on the induced current. 
In essence, the nonlocal effects introduce a quantum capacitance to the field and current variations, which in turn yields significant departure to the local results in the long wavelengths regime. 

Contrary to normal metals, the doping in graphene can be tuned with a vertical electric field \cite{theRiseOfGraphene}.
Hence, the crossover between local to nonlocal electromagnetic response, which occurs when $q_x \sim k_f$, can be tuned by either doping or physical dimension. 
Fig.\,3a and b quantifies the nonlocal effects in nanoribbon arrays under perpendicular light polarization for varying dopings and widths, in terms of the amount of blue-shift in the plasmon resonance, i.e. $\delta\omega_{pl}/\omega_{pl}$. 
The blue-shifts increase inversely with ribbon width and doping, in reasonable agreement with the simple model of Eq.\,\ref{blueshift} as shown. 
Comparison between the simple model and numerical results yield good agreement. 
In Fig.\,3a, the disagreement at small doping is due to finite temperature effects not accounted for in the simple model. 
In Fig.\,3b, the underestimation is due to higher nonlinear terms in the nonlocal conductivity which are again neglected in the simple model. 
Within the experimentally accessible doping and widths as calculated, the blue-shift in the plasmon resonance can be as large as $10\%$, which might explain partly the significant plasmon blue-shifts observed experimentally in narrow ribbons from local electromagnetic results \cite{grapheneDampingPathways}. 
Spatial dispersive effects introduce an in-plane finite quantum capacitance which lead to smaller kinetic inductance and hence faster plasmons.

For parallel polarization case, nonlocality in the optical conductivity leads to an apparent broadening of the Drude peak at terahertz frequencies. 
Denoting the Drude peak's half-width at half-maximum as $\gamma_{dr}$, the nonlocal effects can be quantified through the relative increase in $\gamma_{dr}$, i.e. $\delta\gamma_{dr}/\gamma_{dr}$. 
Fig.\,3c and d quantifies the broadening due to nonlocal effects for varying dopings and widths respectively. 
Since the dispersive part of the longitudinal conductivity includes terms in various powers of $q_x$ and $q_y$ \cite{grapheneSDModels}, nonlocal contribution to the conductivity increases with $1/W^\beta$. 
This is consistent with the increased broadening as $W$ decreases as shown in Fig.\,3d. 
Like the perpendicular polarization case, the nonlocal contribution decreases with doping, however at a much smaller rate. 
In fact, the relative increase in broadening varies very weakly with the chemical potential. 
For the parallel polarization, there is no resonant behavior and the spatial variations of the field are mainly imposed by the ribbon dimensions. 
This is the reason why nonlocal effects show small variations with doping. 
Indeed, the fractional change in the conductivity is independent of $\mu$ (See Eq.\,6 and \cite{grapheneSDModels}). 
With the doping increase and accordingly a more conductive platform for electron traveling, the fields behavior become more independent from nonlocal effects. 
In the limiting case of a perfect electric conductor, the nonlocal effects become negligible.
\begin{figure}
\centering
\includegraphics[width=3.4in]{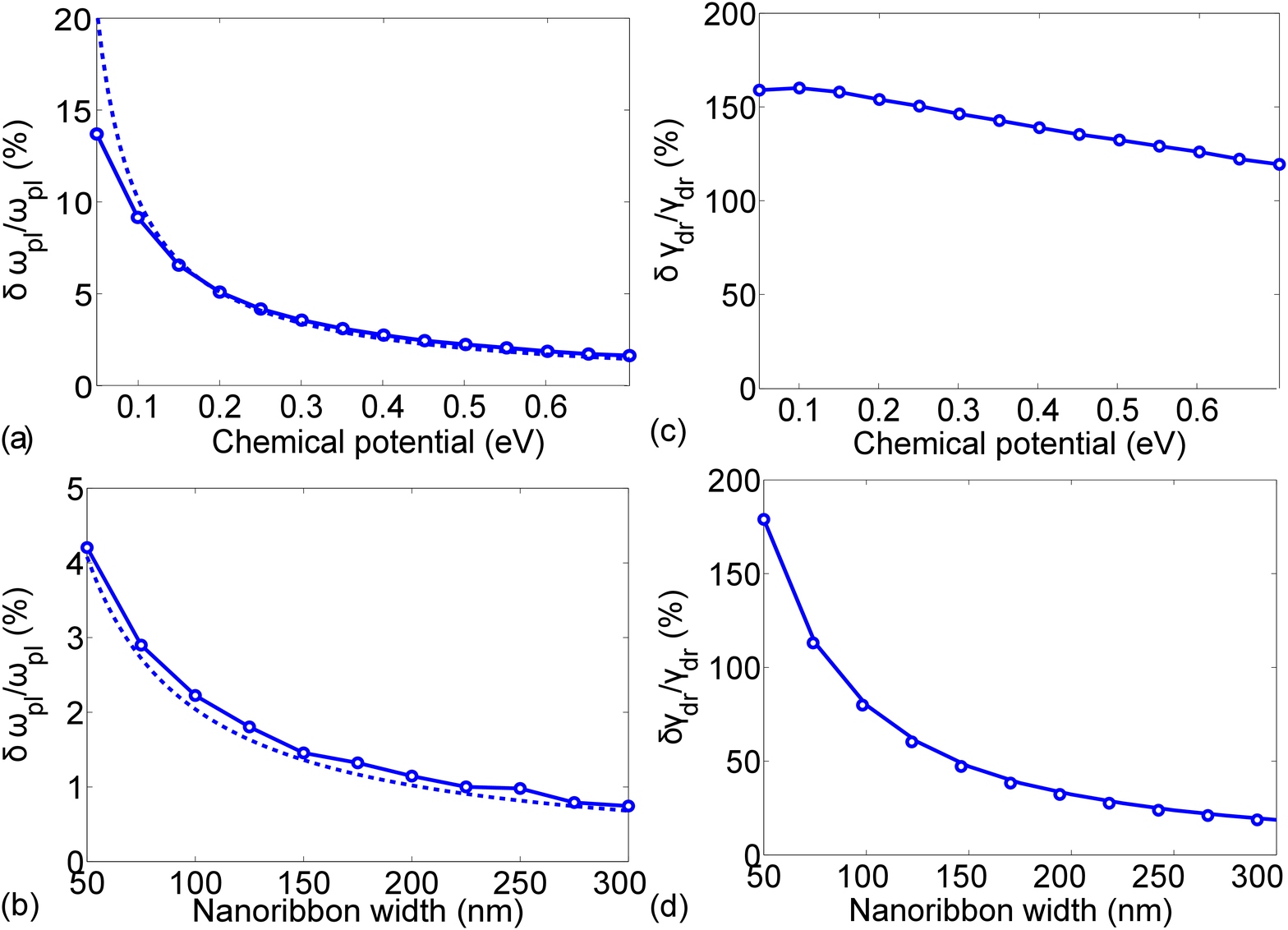}
\caption{Relative shift in the fundamental resonance frequencies, i.e. $\delta\omega_{pl}/\omega_{pl}$, due to nonlocal effects for perpendicularly polarized light transmission across a graphene nano-ribbon array, for (a) varying chemical potential at given width $W=60\,\mathrm{nm}$, and (b) varying widths at given chemical potential of $0.3\,$eV. Similar studies for parallel polarization case, where relative broadening in the Drude conductivity's half-width at half-maximum, i.e. $\delta\gamma_{dr}/\gamma_{dr}$, are displayed in (c) and (d). Dashed lines are model calculations based on Eq.\,7.}
\label{muStudy}
\end{figure}

\emph{Conclusion---}
In this work, we present a modified method of moment approach that solves the Maxwell's equations for a patterned graphene including its general dispersive conductivity. 
We consider the simplest geometry, a periodic array of nanoribbons, where nonlocal effects can be observed. 
For the perpendicular and parallel light polarization, nonlocal effects are manifested as a blue-shift of the plasmon peaks and additional broadening of the Drude peak respectively. 
We quantified how these nonlocal effects depends on graphene's doping and physical dimensions. 
Interpretation of the optical response of graphene nanostructures without accouting for nonlocal effects might lead to inaccurate inference about basic physical parameters, in particular the electronic lifetimes. 
The approach we outlined is able to adapt to arbitrary shapes and configurations. 
In more complicated geometries, one has also the presence of an azimuthal current \cite{grapheneSDModels}, in addition to the longitudinal current studied in this work for ribbons array. 
Our study will serve useful for future investigations of graphene metasurfaces for terahertz and mid-infrared devices.

\emph{Acknowledgement---} We dedicate this work to the memory of Julien Perruisseau-Carrier, who passed away during the preparation of this manuscript.


\end{document}